\documentclass[iop]{emulateapj} 
\usepackage{natbib} 
\usepackage{url} 
\usepackage[usenames,dvipsnames]{color} 
\usepackage{graphicx} 
\usepackage{multirow}
\usepackage{enumitem}
\setlist{noitemsep} 
\newcommand{\frbwG}{$\Delta\nu_\mathrm{GP}/\nu$}
\newcommand{\frbwi}{$\Delta\nu_\mathrm{inst}/\nu$}
\newcommand{\gr}{$\gamma$-ray }
\newcommand{\grs}{$\gamma$-rays }
\slugcomment{version 3.2} 
\shorttitle{A broadband radio study of PSR B1821$-$24A} 
\shortauthors{A.~V.~Bilous et al.}

\begin{document} 
\title{A broadband radio study of the average profile and giant pulses from PSR B1821$-$24A}

\author{A.~V.~Bilous\altaffilmark{1,2}}
\author{T.~T.~Pennucci\altaffilmark{2}}
\author{P.~Demorest\altaffilmark{3}}
\author{S.~M.~Ransom\altaffilmark{4} }

\altaffiltext{1}{Department of Astrophysics/IMAPP,
              Radboud University Nijmegen,
              P.O. Box 9010,
              6500 GL Nijmegen,
              the Netherlands
              \email{a.bilous@science.ru.nl}}
\altaffiltext{2}{Department of Astronomy, University of Virginia, PO
  Box 400325, Charlottesville, VA 22904, USA} 
\altaffiltext{3}{National Radio Astronomy Observatory, P.O. Box O, Socorro, NM 87801, USA}
\altaffiltext{4}{National Radio Astronomy Observatory, 520 Edgemont Road, Charlottesville, VA 22903, USA}

\begin{abstract} 
  We present the results of a wideband (720$-$2400\,MHz) study of PSR~B1821$-$24A (J1824$-$2452A, M28A), 
  an energetic millisecond pulsar visible in radio, X-rays and $\gamma$-rays. In radio, the pulsar has 
  a complex average profile which spans $\gtrsim85$\% of   the spin period and exhibits strong evolution 
  with observing frequency. For the first time we measure phase-resolved polarization properties and 
  spectral indices of radio emission throughout almost all of the on-pulse window. We  synthesize our 
  findings with high-energy information to compare M28A to other known $\gamma$-ray millisecond 
  pulsars and to speculate that M28A's radio emission originates in multiple regions within its magnetosphere 
  (i.e. both in the slot or outer gaps near the light cylinder and at lower altitudes above the polar cap). 
  M28A is one of the handful of pulsars which are known to emit Giant Pulses (GPs) -- short, bright 
  radio pulses of unknown nature. We report a drop in the linear polarization of the average profile in 
  both windows of GP   generation and also a `W'-shaped absorption feature (resembling a double notch), 
  partly overlapping with one of the GP windows.  
   The GPs themselves have broadband spectra consisting of 
  multiple patches with $\Delta \nu/\nu\sim 0.07$. 
  Although our time resolution was not sufficient to resolve the GP structure on the $\mu$s 
  scale, we argue that GPs from this pulsar most closely resemble the GPs from the main pulse of the Crab 
  pulsar, which consist of a series of narrowband nanoshots.
  \end{abstract}


\section{Introduction}

PSR~B1821$-$24A (hereafter M28A) is an isolated 3.05-millisecond pulsar (MSP) in the globular cluster 
Messier~28 \citep{Lyne1987}. The pulsar has an uncommonly large period derivative, $\dot{P}=1.62\times10^{-18}$\,s~s$^{-1}$ 
\citep{Foster1988,Verbiest2009}, about two orders of magnitude larger than typical $\dot{P}$  values for
MSPs. The large observed $\dot{P}$ is a reasonable estimate of the pulsar's magnetic braking: 
based on the measured proper motion of the pulsar and the models of  the gravitational potential inside the 
cluster, \citet[][hereafter J13]{Johnson2013} showed that no more than 14\% of the observed $\dot{P}$ 
can be caused by the accelerated motion in the cluster's gravitational potential and the Shklovskii effect. 

M28A is, in many regards, an extreme member of the MSP population. It is the most energetic MSP known, 
with  an inferred spin-down luminosity ($\dot{E}\propto P^{-3}\dot{P}$) of $2.2\times10^{36}$\,erg~s$^{-1}$. 
M28A has the second-to-largest (after B1937+21) inferred value of the  strength of the magnetic field evaluated 
at the light cylinder ($B_{\mathrm{LC}}\propto P^{-2.5}\dot{P}^{0.5}$, $7.2\times 10^5$~G). The pulsar has known 
irregularities in its spin-down \citep[timing noise; J13,][]{Verbiest2009} and was reported to have a microglitch 
\citep{Cognard2004}.

Pulsed emission from M28A had been detected in radio \citep{Lyne1987}, X-ray \citep{Saito1997} and \grs 
\citep[][J13]{Wu2013}. So far, combining the information from different parts of the electromagnetic 
spectrum has proven to be inconclusive and suggested a complex relationship between the emission regions (J13). 

In radio, M28A has a complex multi-peaked profile, which spans almost the whole pulsar spin period. The 
profile has a high degree of linear polarization and exhibits strong evolution as a function of frequency 
\citep{Ord2004}. Unlike the majority of X-ray-detected MSPs, which show broad thermal X-ray pulsations, 
M28A's X-ray profile consists of two sharp peaks of highly beamed, non-thermal magnetospheric radiation 
\citep{Zavlin2007,Bogdanov2011}. In \grs the pulsar has two broad peaks which are roughly coincident with 
two out of the three main radio peaks. It is hard to classify M28A as a Class I, II or III \gr pulsar 
\citep[with classes defined by the relative phase lag between the radio/\gr profile peaks and presumably 
different radio/$\gamma$-ray emission regions; see ][]{Johnson2014}, although no direct simulation of \gr 
light curves had been performed so far (J13).  

Similar to other pulsars with magnetospheric X-ray emission and comparable values of $B_\mathrm{LC}$ and 
$\dot{E}$, M28A is known to emit giant radio pulses \citep[GPs;][]{Romani2001}. GPs are a rare (known only 
for a handful of pulsars) type of single pulses of unknown origin. GPs are usually distinguished by their 
large brightness temperature (up to $5\times10^{39}$\,K), short duration (ns -- $\mu$s) 
%, high degree of polarization (up to 100\%) 
and power-law energy distribution \citep[see][and references therein]{Knight2006a}.

In this work we analyze an extensive set of broadband, (720$-$920, 1100$-$1900, and 1700$-$2400\,MHz) 
full-Stokes  observations of M28A. The high  signal-to-noise ratio (S/N)  of the accumulated average profiles, 
together with the large fractional bandwidth, allowed us to measure the phase-resolved spectral and polarization properties 
 even of the faint profile components, thus substantially improving upon similar such measurements previously 
made by \citet{Foster1990} and \citet{Yan2011b}.
Such measurements will facilitate constraining 
the location of the radio emission regions and will help multi-wavelength light curve modeling 
\citep{Guillemot2012}. We also collected the largest-to-date sample of M28A's GPs,  which is an order of magnitude 
larger than the sample used in the  most recent previous study \citep[][hereafter K06a]{Knight2006b}. For the first time 
we were able to explore broadband spectra of M28A's GPs and compare them to similar studies of GPs from 
B1937+21 and the Crab pulsar. This is interesting because it is still unclear which properties are common 
 to all GPs, and whether GPs have more than one emission mechanism \citep{Hankins2007}. 

The rest of the paper is organized as follows. After  describing the calibration and correction for propagation 
effects (Section~\ref{sec:obs_calib}), we describe the properties of the average profile (Section~\ref{sec:AP}) 
and the giant pulses (Section~\ref{sec:gps}), comparing the latter to the GPs from the Crab pulsar and 
PSR~B1937+21. In Section~\ref{sec:disc} we compare the properties of M28A to the MSPs from the second 
Fermi catalog \citep{Abdo2013} and speculate on the location of the radio emission regions in the magnetosphere 
of M28A.  A short summary is given in Section~\ref{sec:sum}.

\section{Initial data processing} 
\label{sec:obs_calib} 

M28A was observed with the 100\,m Robert C. Byrd Green Bank Telescope during 21 sessions in 
2010$-$2013. The signal was recorded with the GUPPI\footnote{\url{https://safe.nrao.edu/wiki/bin/view/CICADA/NGNPP}} 
pulsar backend in the coherent dedisperion search mode \citep{DuPlain2008}.
 In this mode, raw voltages were sampled with a time resolution of 1/BW (where BW is the bandwidth). 
Each 512-sample block was Fourier transformed and the signal was coherently
 dedispersed within each frequency channel using the average dispersion measure (DM) for the 
pulsars in the globular cluster M28\footnote{ \url{http://www.naic.edu/$\sim$pfreire/GCpsr.html}}, 
120\,pc\,cm$^{-3}$. After square-law detection, several consecutive Fourier spectra were averaged together 
to meet the disk write speed limits.

Table~\ref{table:obssum} summarizes  some details of the observations. 
Following the standard IEEE (Institute of Electrical and Electronics Engineers) radio frequency naming 
convention\footnote{\url{http://standards.ieee.org/findstds/standard/521-2002.html}}, the observations with the 
central frequencies of  820\,MHz, 1500\,MHz and 2000\,MHz will be hereafter referred to as  UHF, L-band and S-band 
observations. 
In all three bands the time resolution of single-pulse data 
 was $t_\mathrm{res}=10.24$\,$\mu$s and the number of bins in the folded profiles $n_\mathrm{bin}=298$ 
was chosen to match $t_\mathrm{res}$ closely. However,  the UHF single-pulse data were  rewritten with four 
times lower time/frequency resolution shortly after making folded pulse profiles,  and so only these data with 
$t_\mathrm{res}=40.96$\,$\mu$s and a frequency resolution of 1.56\,MHz were available for the UHF part of our GP analysis.

All further data reduction was done using the PSRCHIVE\footnote{http://psrchive.sourceforge.net/} and 
PRESTO\footnote{http://www.cv.nrao.edu/$\sim$sransom/presto/} software packages \citep{Hotan2004,vanStraten2012,Ransom2001}.

The observations presented in this work were a part of the larger dataset analyzed in \citet[][hereafter PDR14]{Pennucci2014} 
and we folded our data with the ephemeris determined from PDR14's dataset. The root-mean-square (RMS) 
deviation of timing residuals, obtained from this ephemeris was 1.35\,$\mu$s, much smaller than the time 
resolution of our data.

Prior to each observation we recorded a pulsed calibration signal, which was used together with standard 
unpolarized flux calibrators (quasars B1442+101 for L-band and 3C190 for S-band and  the UHF observations) 
to correct for the instrumental response of the receiver system.  Polarization calibration for L-band and the 
 UHF observations was conducted using pre-determined Mueller matrix solutions, which described the 
cross-coupling between orthogonal polarizations in the receivers \citep{vanStraten2004}. The Mueller matrix 
was determined using the PSRCHIVE task \texttt{pcm} based on observations of PSR~B0450$+$55 in L-band and 
PSR~B1744$-$21A  in the UHF band.  For S-band polarization calibration we assumed that the feed is ideal and 
composed of two orthogonally polarized receptors. The equivalent flux of the local pulsed calibration signal 
was determined separately in each polarization using observations of the unpolarized source 3C190.  The 
calibration signal measurements were used to balance the gain of each polarization in our S-band observations of 
M28A.  Comparing the polarization of the average profile in the frequency region where L- and S-band overlap 
did not reveal any significant discrepancies between the L- and S-band polarization data. 
In this work we use  the standard PSRCHIVE PSR/IEEE convention for the sign of circular polarization,
described in \citet{vanStraten2010}.

The data in approximately 5\% of the frequency channels  in the UHF band, 25\% of channels in L-band and 20\% 
in S-band were affected by radio frequency interference (RFI) and removed during the calibration (see 
Fig.~1 in PDR14).

\begin{table}
\begin{center} 
\caption{Observing summary. The columns are:  name of the band, its central frequency, bandwidth,
number of channels, time resolution, number of observing sessions per band and the total observing time.\label{table:obssum}}
\begin{tabular}{ccccccc} 
\hline\\ %[0.01cm]
\parbox{0.8cm}{\centering  Band name}&
\parbox{0.8cm}{\centering $\nu_c$ (MHz)} &
\parbox{0.9cm}{\centering BW (MHz)} & 
\parbox{0.9cm}{\centering $N_\mathrm{chan}$} & 
\parbox{1.0cm}{\centering $t_\mathrm{res}$ ($\mu$s)} &  
\parbox{0.8cm}{\centering $N_\mathrm{sess}$} &
\parbox{0.9cm}{\centering Total time (hr)} \\ [0.3cm]
\hline\\
  UHF& 820 & 200 & 512$^\mathrm{a}$ & 10.24\footnote{For the GP analysis, only the data with $t_\mathrm{res}=40.96$\,$\mu$s and 
 $N_\mathrm{chan}=128$ were available.}  & 1 & 2.6\\ [0.1cm]
  L & 1500 & 800\footnote{Some fraction of the band was affected by terrestrial  radio frequency interference, see text for details.} & 512 & 10.24 & 11 & 27.1\\ [0.1cm]
  S & 2000 & 800$^\mathrm{b}$ & 512 & 10.24 & 9 & 14.6\\ [0.1cm]
\hline 
\end{tabular} 
\end{center}
\end{table}

\subsection{Selection of GP candidates}
\label{subsec:gpsearch}

The search for GP candidates was done in the following manner. First, we used PRESTO to dedisperse and 
band-integrate the raw data with the DM from Section~\ref{subsec:DM}. When necessary, we applied an RFI 
mask, calculated from a small subset of data for each session. This resulted in one-dimensional, uncalibrated 
total intensity time series. Then, the candidates were selected from these time series using 
\texttt{single\_pulse\_search.py}  from PRESTO. This routine a) normalizes a chunk of data (setting its mean to 0 and 
RMS to 1); b) convolves the data with a series of boxcar functions of varying width $n$ and a height of 
$1/\sqrt{n}$; c) selects the candidates  from the convolved signal  that have peak values above  a user-specified 
threshold;  d) sifts the duplicate candidates (i.e. those candidates that are above the threshold for different boxcar widths, 
but have the same time of arrival) by comparing the convolved signals corresponding to different boxcar widths 
and selecting the width that gives the largest peak value in the convolved signal. 
Such an optimal width $n_\mathrm{opt}$ will be close to the observed pulse width as seen in the original time-series.

We set the lower limit on the boxcar width to one sample and the upper limit to six samples. The upper limit 
corresponds to the candidate width of 61\,$\mu$s for L- and S-band and 246\,$\mu$s for  the UHF band, which is much larger 
than the expected width of GPs according to K06a. The selection threshold was set  to $\mathrm{SN_{thr}}=7$.

Note that since candidate selection is performed on the convolved signal, this results in an effective 
width-dependent S/N threshold of 7/$\sqrt{n_\mathrm{opt}}$ for the signal with original time resolution. 
The averaging of the signal over $n_\mathrm{opt}$ samples in the convolution reduces the noise variance by the 
same factor, introducing the 1/$\sqrt{n_\mathrm{opt}}$ dependence in the S/N threshold \citep[see details in][]{Bilous2012}. 
 
Knowing the time of arrival of each event, we dedispersed, calibrated and removed the Faraday rotation for 
the corresponding portions of raw data in the same way as for the average profile. About 4\% of candidates 
were discarded since no pulses were revealed during a visual inspection of both the band-averaged and 
frequency-resolved calibrated data. Such candidates were most likely due to short intermittent RFI (missed by
a constant RFI mask, but removed during calibration). 

Note that since some RFI showed intermittency on short (about a minute) timescales, the number of corrupted 
channels varied from one GP to another and did not necessarily coincide with the number of channels removed from 
the time-averaged data from the same observing session.

\subsection{Correction for dispersion  from the interstellar medium  (ISM)}
\label{subsec:DM}

PDR14 measured the  dispersion measure for each of their observing sessions by fitting a two-dimensional (frequency/phase) 
model of the pulse profile to the time-averaged data from each session. Such method allows one to measure the 
variation of DM between the sessions quite precisely (e.g. $6\times10^{-5}$\,pc\,cm$^{-3}$ for our L-band 
sessions in PDR14), but it leaves the absolute value of DM unknown to within some constant value that depends 
on the choice of profile model. Microsecond-long, bright and broadband GPs offer a cross-check on DM determination 
(K06a). For our observing setup, however, the precision of $\mathrm{DM_{GP}}$ obtained by maximizing S/N of a 
giant pulse over the set of trial DMs was limited by the coarse time resolution, which exceeded the expected 
width of GP. In this case, the uncertainty in $\mathrm{DM_{GP}}$ approximately corresponds to the amount of 
DM variation, needed to cause the signal delay of $t_\mathrm{res}$ between the top and the bottom of the band. 
This expected DM uncertainty matched well the standard deviation of the $\mathrm{DM_{GP}}$ values within a 
single observing session (typically, 0.002\,pc\,cm$^{-3}$ for L-band, 0.006\,pc\,cm$^{-3}$ for S- and  the UHF bands, 
two orders of magnitude larger than in PDR14). Nevertheless, in all three bands, in each observing 
session the values of $\mathrm{DM_{GP}}$ were significantly larger than the corresponding $\mathrm{DM_{PDR14}}$ 
for that session and, when dedispersed with $\mathrm{DM_{PDR14}}$, GPs showed visible dispersive delay at the 
lower edge of the band. The low number of pulses per session (especially in S-band, where some sessions had
less than 5 GPs) limited the possibility of refining DM measurements by averaging DMs from individual GPs. 
Thus, for the subsequent analysis we chose not to use $\mathrm{DM_{GP}}$ directly, but rather dedispersed the data with 
$\mathrm{DM_{fold}}=\mathrm{DM_{PDR14}}(t)+const$, where $const=0.011$\,pc\,cm$^{-3}$ was the average difference 
between $\mathrm{DM_{GP}}$ and $\mathrm{DM_{PDR14}}$ for all sessions. 

Phase-resolved measurements of position angle or spectral index in Section~\ref{sec:AP} are somewhat sensitive 
(especially near the edges of profile components) to the value of DM used for folding. To probe the influence 
of the uncertainty in $\mathrm{DM_{fold}}$ on the phase-resolved measurements, we repeated the analysis 
described in Section~\ref{sec:AP} for the data folded with $\mathrm{DM_{fold}} \pm 0.005$\,pc\,cm$^{-3}$. 
Such an estimate of uncertainty in $\mathrm{DM_{fold}}$ corresponded to  the typical standard deviation of 
$\mathrm{DM_{GP}}$ within a single session, averaged between the three observing bands. The discrepancies 
between the values of phase-resolved parameters obtained for different $\mathrm{DM_{fold}}$ were further 
incorporated into the measurement uncertainties (see Fig.~\ref{fig:stat}(d$-$f)).

\subsection{Correction for Faraday rotation}
\label{subsec:RM}

 The propagation of a linearly polarized wave through the magnetized plasma in the interstellar medium causes the wave's position 
angle (PA) to rotate by an angle proportional to the square of its wavelength. The coefficient 
of the proportionality,  called the rotation measure (RM), was estimated for each session by fitting the  function 
$\mathrm{PA}=\mathrm{RM}\times(c/\nu)^2 + const$ to the position angle of emission  in the two 
phase windows containing the brightest and most polarized profile components \citep[P1 and P2, following 
the convention in ][ also see Fig.~\ref{fig:stat}(a)]{Backer1997}. The measured values of RM were a sum of 
interstellar $\mathrm{RM_{IS}}$ and a contribution from Earth's ionosphere $\mathrm{RM_{iono}}$, with the latter 
depending on the total electron content along the  line of sight and the orientation of the  line of sight with 
respect to the Earth's magnetic field. We estimated $\mathrm{RM_{iono}}$ with the 
\texttt{ionFR}\footnote{\url{http://sourceforge.net/projects/ionfr/}} software \citep{Sotomayor2013}, which uses 
the International Geomagnetic Reference Field (IGRF11) and global ionospheric maps to predict $\mathrm{RM_{iono}}$ 
along a given  line of sight at a specific geographic location. For our set of parameters, $\mathrm{RM_{iono}}$ 
varied from 1 to 6 rad~m$^{-2}$, and changed by 0.2$-$1.5 rad~m$^{-2}$ during an observing session. 

The typical uncertainty of a single-epoch measurement of $\mathrm{RM_{IS}}=\mathrm{RM}-\mathrm{RM_{iono}}$ was 
about 1\,rad~m$^{-2}$. The values of $\mathrm{RM_{IS}}$ from 21 observing sessions did not show any apparent 
secular variation with time, being randomly scattered around  an average value of $82.5$\,rad~m$^{-2}$ with  an RMS of 
2.1\,rad~m$^{-2}$. The most recent value for $\mathrm{RM_{IS}}=77.8\pm0.6$\,rad~m$^{-2}$, obtained by \citet{Yan2011a}, 
is somewhat lower, but still in agreement with our measurement.

\subsection{Scintillation and scattering}
\label{subsec:scat}

 The giant pulses that we detected in the UHF band and in some of the L-band sessions displayed a typical 
fast-rise/exponential-decay shape.  This is indicative of the GPs having traversed an inhomogeneous ISM (Fig.\,\ref{fig:GPs}), 
and it allowed us to estimate the scattering time $\tau_\mathrm{sc}$ for these sessions.
For each session,  we constructed an average GP by summing the GPs aligned by the phase of half-maximum intensity on the rising edge 
and then fit one-sided exponential to the trailing side of the average GP. In our single  UHF epoch, 
$\tau_\mathrm{sc}=100\pm 40$\,$\mu$s. In L-band, $\tau_\mathrm{sc}$ varied between sessions, from being virtually 
undetectable even at the lower edge of the band, $\tau_\mathrm{sc}(1174.5$\,MHz$)<10$\,$\mu$s, to being as large as 
$\tau_\mathrm{sc}(1174.5$\,MHz$)=25\pm 8$\,$\mu$s. The sessions which were separated by about a month or less 
usually had similar scattering time. The variability of $\tau_\mathrm{sc}$ was previously noted by K06a and our
measurements are consistent with $\tau_\mathrm{sc}$ of 24.5\,$\mu$s measured at 1\,GHz by \citet{Foster1990}.

If $\tau_\mathrm{sc}(1200$\,MHz$)\approx15$\,$\mu$s and the spectrum of turbulence is Kolmogorov, then the 
decorrelation bandwidth $\delta f \approx 1/(2\pi \tau_{\mathrm{sc}})$ ranges from 1\,kHz at 720\,MHz to 220\,kHz 
at 2400\,MHz,  which is much smaller than our frequency resolution at all observing frequencies (390\,kHz  in the UHF band 
and 1.56\,MHz in L- and S-band). Since many scintles are averaged within one frequency channel, we do not see any 
signs of diffractive scintillation in the observed spectrum of M28A's emission. However, we record the variation 
of the pulsar's flux between observing sessions. The modulation index of such variation ($m=\sigma_I/\langle I\rangle$, 
where $\sigma_I$ is the standard deviation of the observed flux densities and $\langle I\rangle$ is their mean) was 
0.11 for  the eleven L-band observations and 0.17 for  the nine S-band sessions. It is interesting to compare the observed 
flux modulation with the prediction for refractive scintillation (RISS). The predicted characteristic timescale 
of refractive scintillation changes from 50 days at 1100\,MHz to 10 days at 2400\,MHz, assuming the pulsar's 
transverse velocity to be 200\,km~s$^{-1}$ (J13). The expected modulation index for RISS is given by the decorrelation 
bandwidth of diffractive scintillation at a given frequency, $m_\mathrm{RISS}=(\delta f / f)^{1/6}$ \citep{Rickett1996}.
For the center frequencies of L- and S-bands, $m_\mathrm{RISS}$ is expected to be equal to $0.19$ and $0.22$  
respectively, roughly consistent with the observed flux modulation. 

\section{Average profile} 
\label{sec:AP}

\begin{figure} 
\centering 
\includegraphics[scale=0.97]{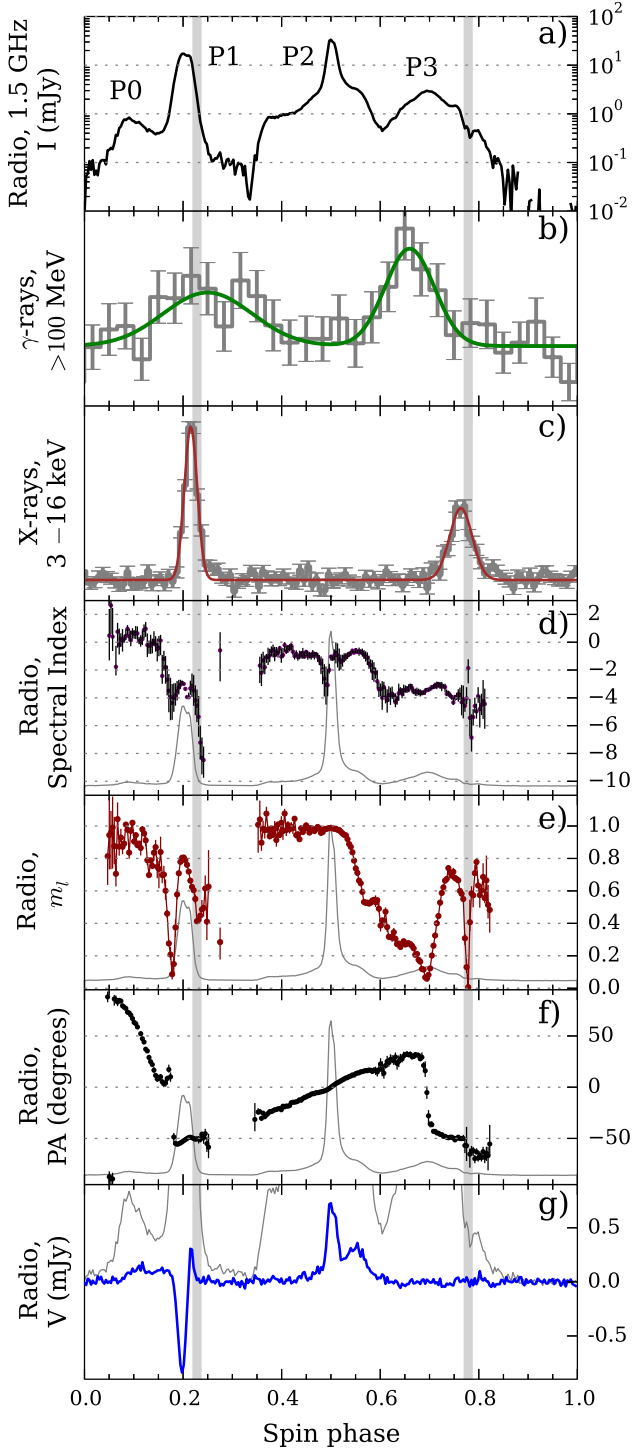}
\caption{Phase-resolved properties of M28A's average profile in L-band (1100$-$1900\,MHz) together with the X-ray and 
\gr profiles from J13 (see text for details on the alignment). \textit{(a)} total intensity of the radio emission $I$ 
on a log scale; \textit{(b,c)} \gr and X-ray profiles together with their respective models (J13); \textit{(d,e,f)} 
phase-resolved spectral indices, fractional linear polarization and relative position angle of the radio emission; 
\textit{(g)} circular polarization $V$ with the total intensity profile \textit{(grey)} overplotted on the same scale. In subplots 
\textit{(d-f)}, the total intensity profile from  subplot \textit{(a)} was overplotted for reference with the intensity 
on the linear scale. The grey-shaded areas mark the regions with GP emission. 
}
\label{fig:stat} 
\end{figure}

Fig.~\ref{fig:stat}(a) shows M28A's average profile, obtained by accumulating the signal from 27 hours of observations 
at L-band. The peak of the brightest component at 1500\,MHz was  aligned to phase 0.5, and the components  near phases 
0.2, 0.5 and 0.7 are labeled P1/P2/P3, following the convention  of \citet{Backer1997}. Additionally, we used ``P0'' 
to label the faint precursor to component P1  (this precursor also appears in profiles
presented by \citet{Yan2011b} and J13). For the complete, frequency-resolved representation of M28A's total 
intensity profiles between 720 and 2400 MHz we refer the reader to Fig.~1 in PDR14.

One of the early works on M28A suggested that this pulsar may have two distinct emission modes. According to 
\citet{Backer1997}, the peak intensity of P2  was three times smaller than the peak intensity of P1 for one third of 
their observing time at 1395\,MHz (so-called ``abnormal mode''). For the rest of the time P2 was about two times 
brighter than P1 (``normal mode''). At the same time, the peak ratio at 800\,MHz did not show any temporal variations 
down to 25\% accuracy. \citet{Romani2001} reported that in their observations at 1518\,MHz, P2's relative flux (with 
respect to P1 and P3)  decreased  by 25\% during the first hour of observations and stabilized after that. Our data 
do not show any signs of such behavior.  The peak intensity ratio P2/P1, estimated by averaging 5 minutes of 
 band-averaged data, fluctuated randomly within 6\%  in UHF and L-band and within 12\% in S-band. Judging by the standard 
deviation of the signal in the off-pulse windows, in all three bands such amount of fluctuation can be attributed to 
the influence of the background noise. It is worth noting that the ``abnormal'' mode, observed by \citet{Backer1997}
could, in principle, be caused by instrumental error. The average profile of M28A exhibits very high levels of linear
polarization with the position angle changing significantly between profile components (Fig.~\ref{fig:stat}(e$-$f)). 
If the signal in one of the linear polarizations was for some reason not recorded, then the observed shape of the average 
profile could be dramatically distorted. We simulated a series of such ``abnormal'' average profiles by trying different 
orientations of the Q/U basis on the sky, with the U component of the Stokes vector being subsequently set to zero. For 
some of the orientations, the resulting profile resembled the ``abnormal mode'' described in \citet{Backer1997}, with 
intensity of P2 being about three times smaller than the peak intensity of P1.

The logarithmic scale of the ordinate axis in Fig.~\ref{fig:stat}(a) allows very faint profile components to be 
distinguished (see also the linear-scale zoom-in on Fig.~\ref{fig:stat}(g)). Interestingly, the average profile occupies 
almost all ($\gtrsim 85$\%) of the  spin period, with the only apparent noise window falling at phases 0.882$-$1.000 
(hereafter defined as  the ``off-pulse window''). Even between P1 and P2 there is radio emission at 0.1\,mJy level (with 
 a significance of $5\sigma$, where $\sigma=0.02$\,mJy is the standard deviation of the data within the off-pulse window).

 Overall, the shape of the total intensity profile agrees well with profiles at similar frequencies
presented by J13 and \citet{Yan2011b}. Similar to J13, we record a small dip at the peak of component P1. 
In the work of \citet{Yan2011b}, this dip seems to be washed-out by the intra-channel dispersive smearing in their data.
According to Table~1 from their work, the dispersive smearing
corresponded to about 0.016 spin periods, commensurate with the width of the dip in our data.

Fig.~\ref{fig:stat}(b$-$c) show M28A's X-ray and \gr average profiles together with their models. Both the profiles 
and the model parameters were taken from J13. We aligned the high-energy profiles with our radio data by noting that 
the radio profile from J13, produced with the same ephemeris as  that used for the high-energy light curves, has the peak of P1 
 at phase 0, whereas in our phase convention the peak of P1 falls at phase 0.2.  The grey-shaded areas 
indicate regions of GP emission (Section~\ref{subsec:gp_phase}) in all subplots of Fig.~\ref{fig:stat}.

The rest of the subplots in Fig.~\ref{fig:stat} feature the phase-resolved measurements of spectral indices and 
polarization of M28A's emission. We show the results obtained from L-band only. The average profile in S-band yields 
similar results, but with worse S/N, and the emission from  the UHF band is scattered by an amount approximately 
ten times larger than the 10.24\,$\mu$s time resolution, making phase-resolved fits unreasonable.

Phase-resolved spectral indices of M28A's radio emission in L-band are shown on Fig.~\ref{fig:stat}(d), with the 
average profile overplotted for reference\footnote{The average profile is identical to the one on Fig.~\ref{fig:stat}(a), 
but with intensity on a linear scale.}. The indices were measured by fitting a power-law function $I=I_0(\nu/\nu_0)^s$ 
(where $\nu_0=1500$\,MHz and $\nu$ goes from 1100 to 1900\,MHz) to the total intensity in each phase bin\footnote{Another 
approach to measuring spectral indices is given in PDR14, where they approximate the profile with the sum of several 
Gaussian components and fit a power-law frequency dependence to the position, width and the height of each component.}. 
As is expected for a pulsar with strong profile evolution, $s(\phi)$ show considerable amount of variation across 
the on-pulse windows. It must be noted, though, that the apparent behavior of the phase-resolved spectral indices 
may be influenced by the frequency-dependent variation of components' position and width. For the former, neither 
visual inspection nor the analysis in PDR14 show any significant changes in the position of the main components 
within each separate band. However, profile components become narrower with increasing frequency, which is causing 
the steepening of the spectral indices at the component edges. Also, the behavior of $s(\phi)$ in any regions with large 
$|dI/d\phi|$ is sensitive to the DM used for folding, which is reflected by the larger errorbars in these regions 
(Section~\ref{subsec:DM}). Scattering, with its steep dependence on observing frequency, can also contribute to the 
steepening of the measured spectral index at the trailing edge of a sharp profile component.

\begin{table}
\begin{center} 
\caption{Spectral indices of emission, averaged within the individual components in the 1100$-$2400\,MHz frequency range.
The data from L- and S-band were combined. P0 stands for the faint precursor to component P1.\label{table:spind}}
\begin{tabular}{rcc} 
\hline\\ %[0.01cm]
& Phase window &  Spectral index
 \\ [0.3cm]
\hline\\
P0 & $0.05-0.15$ & ~~$0.38\pm0.1$  \\ [0.1cm]
P1 & $0.15-0.25$ & $-3.50\pm0.04$ \\ [0.1cm]
P2 & $0.35-0.60$ & $-1.40\pm0.03$  \\ [0.1cm]
P3 & $0.60-0.85$ & $-3.61\pm0.05$  \\ [0.1cm]
\hline 
\end{tabular} 
\end{center}
\end{table}

Except for the steepening in the regions with large $|dI/d\phi|$, the values of $s(\phi)$ stay roughly similar within 
four broad phase windows corresponding to P0, P1, P2 and P3. In order to compare the spectral indices more readily to 
the information available for other MSPs (Section~\ref{sec:disc}), we obtained the spectral indices of emission, 
averaged within each of the four regions. For this fit, we combined the data from L- and S-bands, but omitted the 
 UHF session, since the scattering time in that band corresponded to  a significant (30\%) fraction of the 
width of components P0 or P1. The results are presented in Table~\ref{table:spind}.

 The average profile of M28A is very strongly linearly polarized, and has only a small degree of circular polarization. In this 
work we discuss only relative position angles of the linearly polarized emission, setting PA($\phi=0.5$) to 0. Generally, both 
the amount of linear polarization and the relative PA showed the same phase-resolved behavior in all three bands. However, the 
S/N of the signal was lower in S-band and any rapid variation of $m_l(\phi)$ and $\mathrm{PA}(\phi)$ were washed out by 
scattering  in the UHF band.  In L-band, the shape of the PA curve is similar to the one from 
\citet{Yan2011b}\footnote{ Note that \citet{Yan2011b} report absolute position angles and their measurements cover less 
of the on-pulse window.}.

Both P0 and most of the component P2 are almost completely linearly polarized, with opposite signs of PA gradient.
Components P1 and P3 are less polarized, with $m_l<0.8$. There are two large jumps in PA: one by $\Delta\mathrm{PA}=50^\circ$ 
right before the start of P1 and another one at the peak of P3 ($\Delta\mathrm{PA}\approx90^\circ$). Both jumps are 
accompanied by drops in $m_l$. Interestingly, there is also a drop in $m_l$ in both regions of GP emission: at the 
trailing edge of P1 $m_l$ falls to 0.4 and at the trailing edge of P3 $m_l$ is almost 0. 

\begin{figure}
\centering 
\includegraphics[scale=0.85]{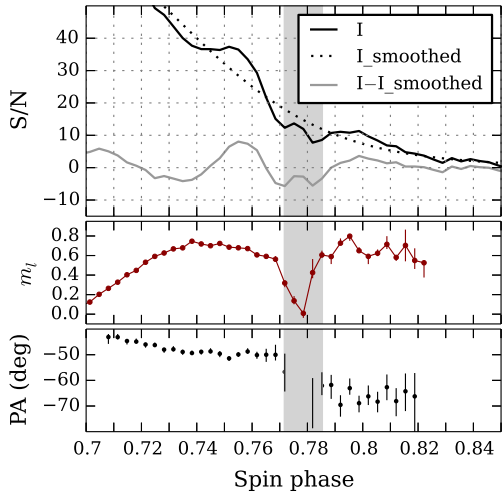}
\caption{\textit{From top to bottom}: zoom-in on the total intensity (in units of S/N), fractional linear polarization, and PA
on a trailing edge of component P3. 
The `W'-shaped feature at the phase 0.777 resembles
a double notch,  which is an absorption feature observed previously in a few other pulsars  that can
potentially  provide clues to the microphysics of pulsar emission \citep{Dyks2012}. 
The grey-shaded area marks the region where GPs occur.  
}
\label{fig:notch} 
\end{figure}

The exceptionally high S/N of the average profile in L-band allowed us to detect an interesting feature on the 
trailing edge of component P3 (Fig.~\ref{fig:notch}). As was pointed out by J.~Dyks (private communication), a 
`W'-shaped dip in the total intensity profile around phase 0.777 resembles the so-called ``double notches'', 
observed previously in the average profiles of normal pulsars B0950+08, B1919+10 and the MSP J0437$-$4715 
\citep[][and references therein]{Dyks2012}. Double notches were proposed to be a result of a double eclipse 
in the pulsar magnetosphere \citep{Wright2004} or a representation of the shape of a microscopic beam of 
emitted coherent radiation \citep{Dyks2007,Dyks2012}.

We compared the properties of M28A's dips with the properties of double notches in the literature. The depth of 
M28A's feature (the drop of the total intensity flux $I$ in the minima with respect to the flux right outside 
the feature) is about 30\%, assuming that the phase window $0.882<\phi<1.000$ does not contain pulsar emission. 
This agrees with the depth of double notches  \citep[20\%$-$50\%,][]{Dyks2012}. For the previously observed double 
notches, their local minima approached each other with  increasing frequency as $\nu^{-0.5}$. Unfortunately, we 
were unable to explore the frequency-dependent dip separation for M28A, since the combination of  a steep spectrum 
of the underlying component and the increasing role of scattering at lower frequencies make M28A's feature visible 
only in a limited frequency range (between 1100 and 1600\,MHz). Interestingly, unlike observed double notches in 
other pulsars \citep{Dyks2007}, the double feature in M28A coincides with changes in polarization behavior: a drop 
in $m_l$ and the small, $15^\circ$ jump in PA. It is unclear whether these changes are connected to the dips or to 
the region of GP generation, which is partially overlapping with the feature.

Alternatively, the bump around phase 0.777 can be interpreted as being due to the extra flux brought by GPs. The 
GPs detected in L-band at the trailing edge of P3 contribute only 0.006\,mJy to the average pulse flux density in 
the corresponding phase window ( an additional S/N of 0.3  in Fig.~\ref{fig:notch}). Although this value is much smaller 
than the height of the bump, the actual GP contribution is unknown, since we have no information about the energy 
distribution of GPs below our detection threshold.

The question of whether this feature in M28A's profile is a double notch still remains open. In any case, the properties 
of the feature are worth exploring and may provide some clues to conditions in M28A's magnetosphere.

\section{Giant pulses} 
\label{sec:gps} 

The nature of GPs is far from being understood and even the precise definition of GP has not been established yet.
In this work, we will define GPs as individual pulses which have power-law energy distribution and either narrow widths 
($<10$\,$\mu$s) or substructure on nanosecond timescales \citep[see discussion in][]{Knight2006a}.

Our time resolution was not  suffucient to resolve the temporal structure of the recorded pulses.  However, 
all of the pulses detected in our study were found within the same phase regions where GPs were previously found by \citet{Romani2001} 
and K06a. Also, the energy distribution of the detected pulses in our study was the same as that of
the $\mu$s-long GPs from the fine time resolution study of K06a. Therefore, we conclude that our pulses 
are GPs in the sense of the aforementioned definition.

In total, we have collected 476 GPs with S/N$>$7 on the band-averaged, calibrated total intensity data. The number 
of observed pulses per band and window of occurrence is given in Table~\ref{table:gpsum}. 
 Our sample is almost 20 times larger than that in \citet{Knight2006b}
and the pulses in our work were recorded within 10 times larger frequency band. Thus, we can make better
measurements of the pulse energy distribution and for the first time probe the instantaneous broadband spectra of a large number of
individual GPs.

\begin{table}
\begin{center} 
\caption{Number of GPs detected in each band in each of  the two phase windows falling at the trailing edges of components 
P1 and P3.\label{table:gpsum}}
\begin{tabular}{cll} 
\hline\\ %[0.01cm]
\parbox{1.1cm}{Band} &
\parbox{1.0cm}{P1} & 
\parbox{1.1cm}{P3} \\ [0.3cm]
\hline\\
  UHF & 27 & 12 \\ [0.1cm]
 L & 273 & 81 \\ [0.1cm]
 S & 64 & 19 \\ [0.1cm]
\hline 
\end{tabular} 
\end{center}
\end{table}

\subsection{Phase of occurrence}
\label{subsec:gp_phase}

The leading edge of GPs detected in L- and S-band fell inside a phase window with a width of 0.013 (corresponding to 
four phase bins or 41\,$\mu$s) starting at phases 0.221 and 0.772.  In the UHF band, the GP window was also four bins wide, but 
four times longer due to having four times lower time resolution. K06a, based on a much smaller number of  GPs detected with better 
time resolution, reported the width of the P1 GP window to be 0.006  spin periods, or 18\,$\mu$s. 
 This indicates that our width of GP window may be overestimated, most probably because of the coarse $t_\mathrm{res}$
and, to some extent, the imperfect DM corrections (see Section~\ref{subsec:DM}). 

Both GP windows are within the on-pulse phase range of X-ray profile components (Fig.~\ref{fig:stat}(b)). However, for both 
P1 and P3 GPs we noticed that the peaks of  the X-ray components from J13 are 0.013/0.015  spin periods ahead of the centers 
of GP windows. Although this is in qualitative agreement with the slight misalignment between the peaks of  the X-ray profile 
components and the centers of GP windows presented in Fig.~1 of K06a, the magnitude of the lag must be taken with caution 
until measured by direct comparison of radio and X-ray data folded with the same ephemeris.

\begin{figure}
\centering 
\includegraphics[scale=0.8]{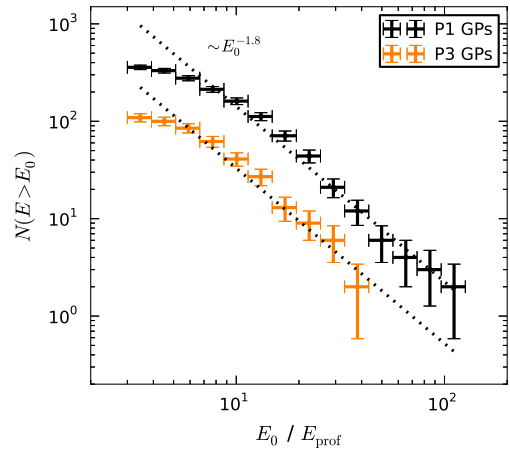}
\caption{Cumulative energy distributions for GPs coming from the trailing edges of components P1 and P3 in all three 
observing bands. The errorbars correspond to the $\sqrt{N}$ Poisson uncertainties and do not account for the 
under-representation of low-energy GPs in our sample. The low-energy end of distribution is affected by a number 
of selection effects, described in the text. 
}
\label{fig:Edistr} 
\end{figure}

\begin{figure*}
\centering 
\includegraphics[scale=0.75]{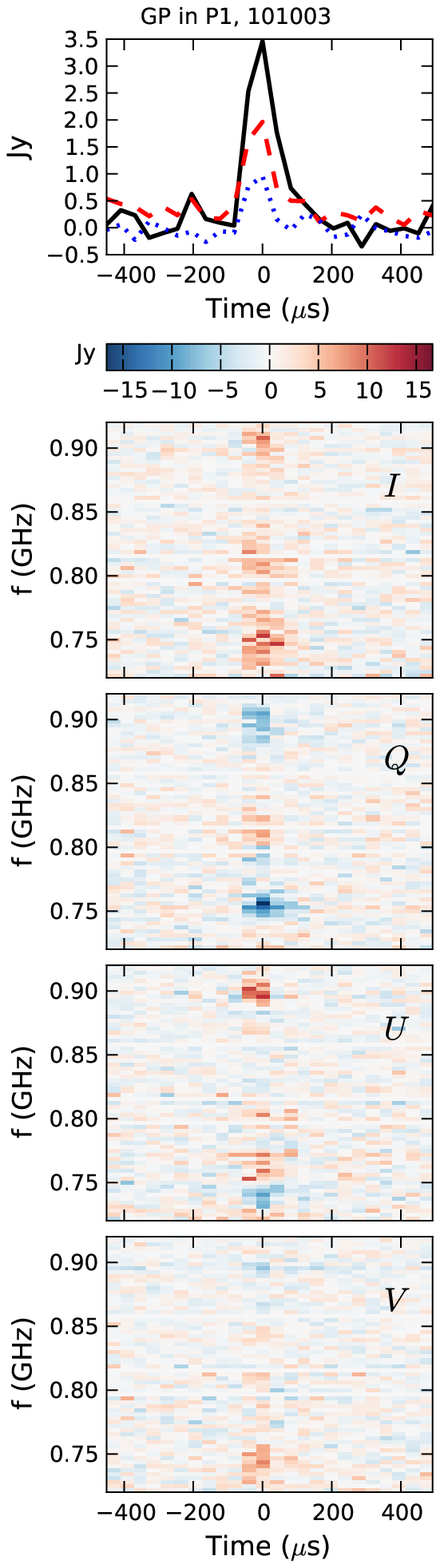}\includegraphics[scale=0.75]{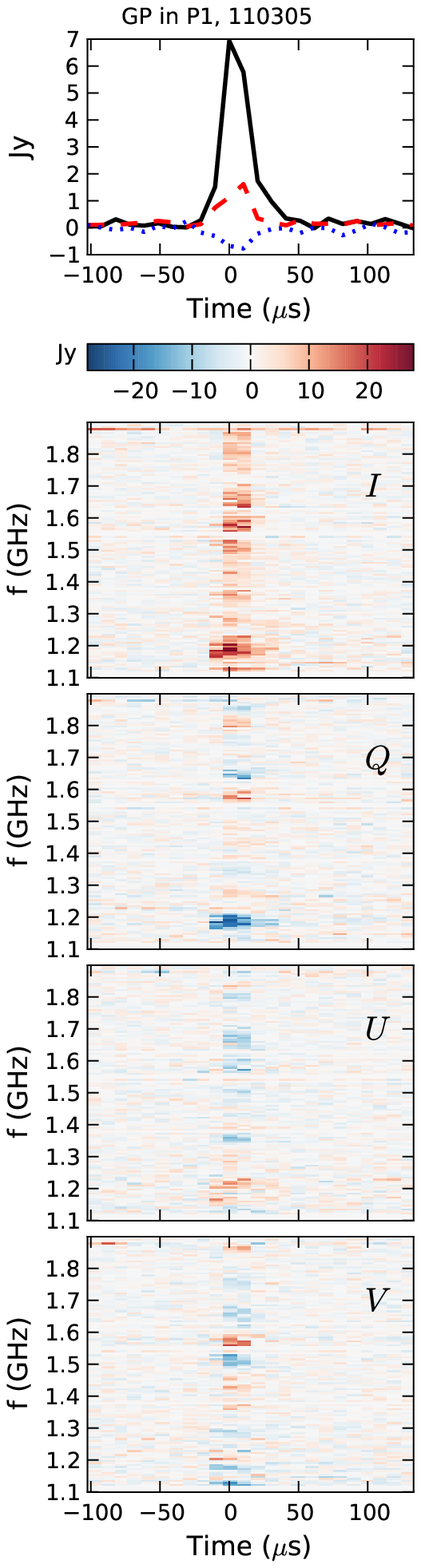}\includegraphics[scale=0.75]{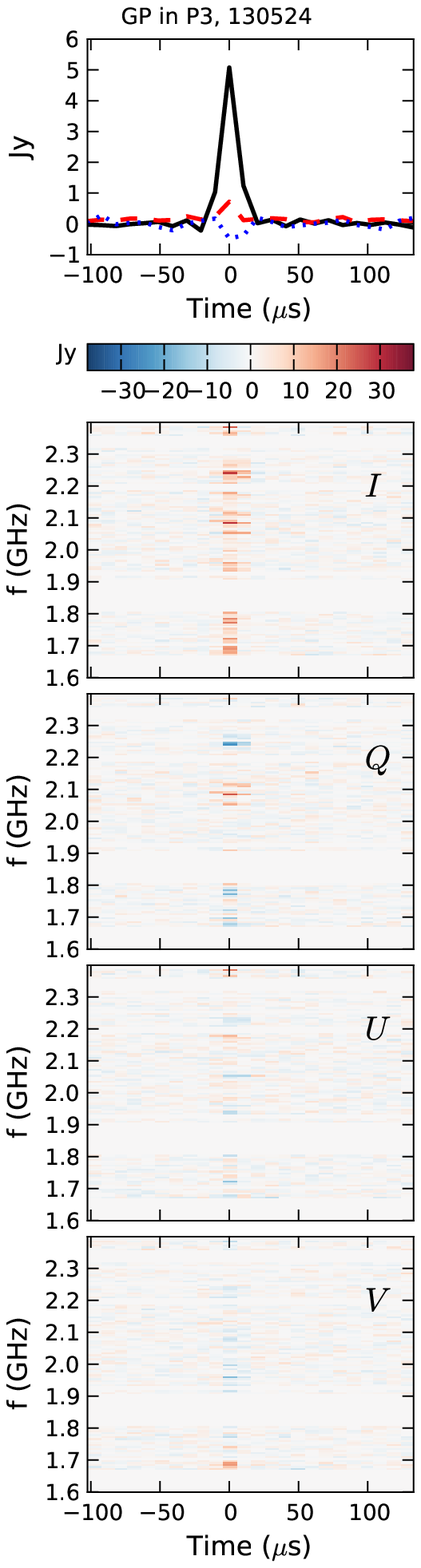}
\caption{Examples of GPs from the three observed bands in four Stokes components.  The phase window of arrival (trailing 
  edge of component P1 or P3) is labeled on top together with  the observing date in \textit{yymmdd} format.  The top row 
  features band-integrated emission, with total intensity $I$ (\textit{solid black line}), linear polarisation 
  $\sqrt{Q^2+U^2}$ (\textit{red dashed line}) and circular polarization $V$ (\textit{dotted blue line}). The rest 
   of each column shows  the spectra of GPs from the top row in each of the four Stokes components.  The GPs are mostly unresolved 
  in time and have exponential scattering tails at lower frequencies (at the bottom of  the UHF and L-band observation). 
  Note that both the time and frequency resolutions of  the UHF data are four times lower than in L- and S-bands. The decorrelation bandwidth 
  of scintillation due to multipath propagation in the ISM is much smaller than the width of the frequency channels. 
  Thus, the observed frequency structure is an intrinsic property of M28A's GPs. The polarization was corrected for 
  Faraday rotation  from the interstellar medium and Earth's ionosphere (see Section~\ref{subsec:RM}). Within  the GPs, the
  position angle of linear polarization and the sign of circular polarization varies from patch to patch. Therefore,  
  band-integrated GPs have a smaller degree of linear and/or circular polarization.}
\label{fig:GPs} 
\end{figure*}

\subsection{Pulse widths}
\label{subsec:gp_width}

While searching for GPs, we defined the width of each GP candidate as the width of the boxcar function which 
 led to the largest peak value in the convolved signal (Section~\ref{subsec:gpsearch}). 
Thus, our measured width of GPs could assume only integer multiples 
of the sampling time $t_\mathrm{res}$. In S-band and some of the sessions in L-band all GPs had widths of 1 or 2 
samples (10$-$20\,$\mu$s), indicating that the pulses were unresolved (2-sample GPs can occur when the pulse falls 
between the phase bins). This agrees with K06a, who measured the FWHM of GPs to be between 750\,ns and 6\,$\mu$s 
at 1341\,MHz. For some of the sessions in L-band, the GPs were wider and had a clear scattering tail. GPs 
 in the UHF band had widths of 1$-$4 samples (41$-$164\,$\mu$s) and strong GPs showed clear signs of scattering.

\subsection{Energy distribution}

The cumulative energy distributions for P1 and P3 GPs are shown in Fig.~\ref{fig:Edistr}. In order to reduce the 
influence of refractive scintillation, in each session GP energies were normalized by the energy of the average 
profile on that day, $E_\mathrm{prof}$. Distributions from different bands did not show any substantial difference,
so we combined all GPs together. We believe that the flattening of the distribution at the low-energy side is due 
to the under-representation of the pulses close to the S/N threshold, caused by: a) the width-dependent selection 
threshold (see Section~\ref{subsec:gpsearch}); b) the variation of pulsar flux between the sessions due to refractive 
scintillation (same S/N threshold corresponds to different $E/E_\mathrm{prof}$); c) slow variation of background noise within 
each session due to an increased atmospheric path length when the pulsar was approaching the horizon.

Both P1 and P3 GPs have  a power-law energy distribution with  an index of $-1.8\pm0.3$\footnote{ The uncertainty was obtained by 
varying the lowest energy threshold used in the power-law fit.}, in agreement with $-1.6$ from K06a. The strongest pulses in all 
three bands had energies about 100 times larger than the typical energy of the average profile on that session. Since 
GPs occupy less than 1\% of pulsar phase, the mean flux during such pulses exceed the average pulsar flux by $>10^4$ times.

\begin{figure*}
\centering 
\includegraphics[scale=0.8]{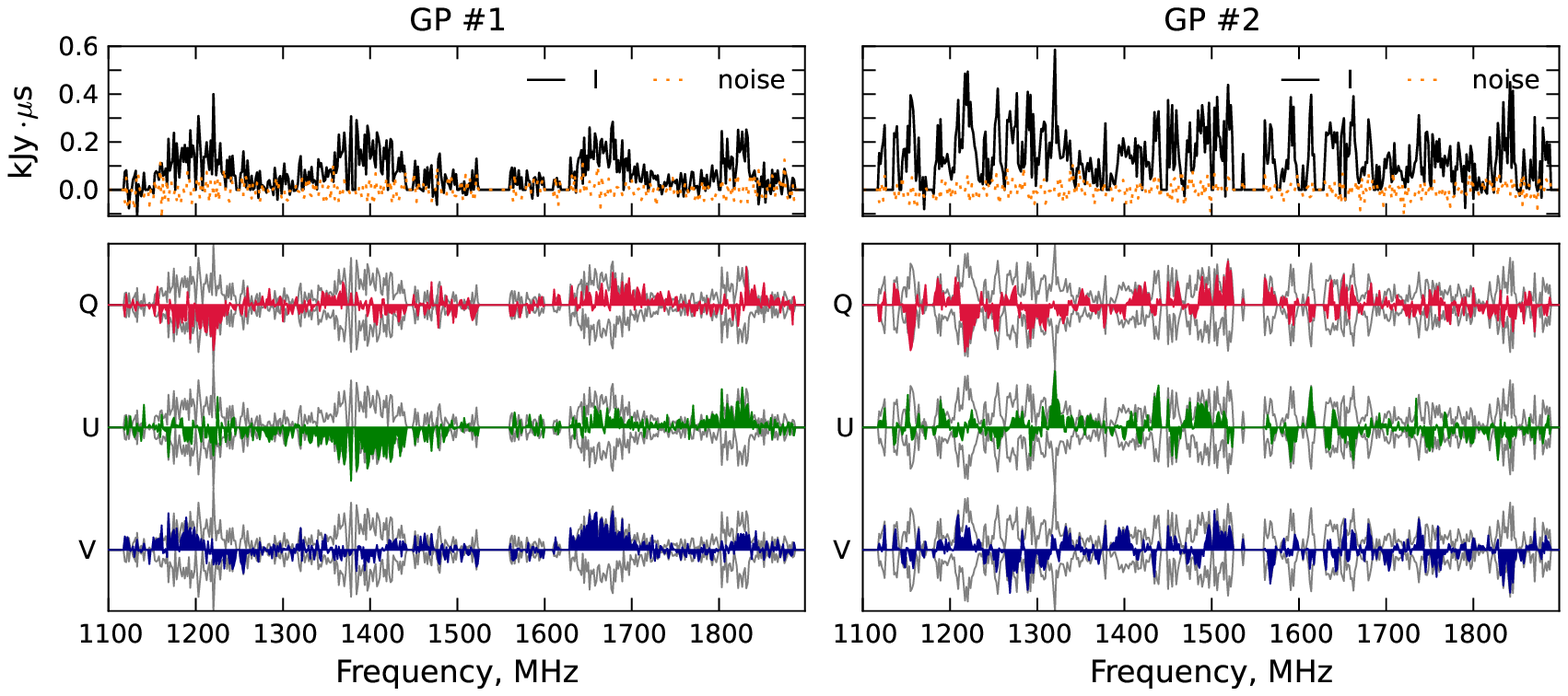} 
\caption{Spectra of two P1 GPs recorded 20 minutes apart from each other. \textit{Top:}  the total intensity of GP $I(\nu)$ 
 (black line) and the level of noise outside the pulse  (orange dots). The signal in some of the channels was set to zero 
because of RFI. \textit{Bottom:}  the three other components of the Stokes vector for the same pulses. $I(\nu)$ and $-I(\nu)$ 
 are plotted in grey together with each polarized component for reference. The GP on the left shows visible $\sim100$\,MHz 
clumps with polarization being constant or changing slowly within each clump. The GP on the right has  a more complex frequency 
structure, presumably consisting of $\approx10$\,MHz clumps with similar polarization behavior. 
}
\label{fig:patches} 
\end{figure*} 

\begin{figure}
\centering 
\includegraphics[scale=0.75]{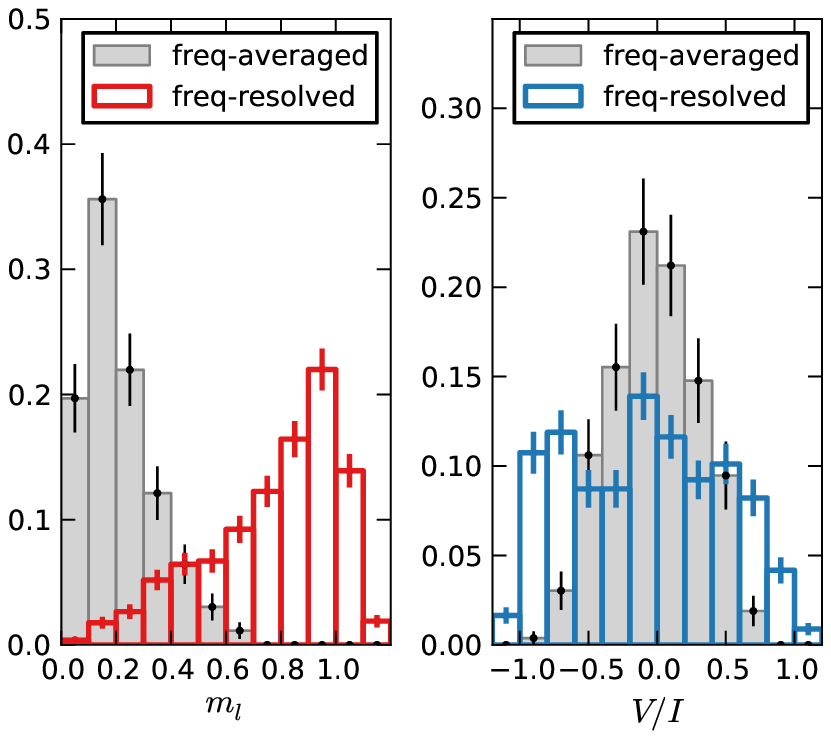}
\caption{Fractional polarization of GPs: linear (left) and circular (right). Filled bars show  the band-integrated fractional 
polarization for all GPs with $\mathrm{S/N>15}$. Hollow bars show  the polarization per 1.56\,MHz frequency channel for all 
channels where  the S/N of the total intensity signal was $>15$. Histograms are normalized by the sum of bar heights and 
Poisson errors are shown on top of each bar. Apparent fractional polarization with magnitude above 100\% is due 
to the influence of noise (see text for details). Individual GP patches tend to be more polarized, but the PA varies 
from patch to patch which causes gradual washing-out of the polarization across the band.  
Depending on the intrinsic GP width (which is unresolved in our single-pulse data), the presented polarization measurements may  
be biased by small-number statistics, and thus may not reflect the intrinsic
polarization properties of the GP emission mechanism (see text for details).
}
\label{fig:GP_pln} 
\end{figure}

\subsection{Spectra and polarization}

Despite having observed in a quite narrow band (64\,MHz), K06a noticed some prominent frequency variability in the 
spectra of individual GPs, which led the authors to suggest that GP spectra consist of  a series of narrow-band patches. 
 Having observed with a much larger bandwidth, we can clearly see patchy structure  in the spectra of individual GPs 
(Fig.~\ref{fig:GPs}).  The patches occur more or less uniformly across the observing bands and they cannot be caused by 
scintillation due to multipath propagation in the ISM -- the decorrelation bandwidth is much smaller than the width 
of any single frequency channel at all our observing frequencies.

After visual examination, we failed to notice any difference between the spectra of P1 and P3 GPs. However, interestingly, 
the characteristic size of the frequency structure was noted to vary from pulse to pulse. We explored the size of the 
frequency structure by constructing auto-correlation functions (ACFs) for the one-dimensional spectra of bright pulses 
with $\mathrm{S/N}>20$. Pulses were averaged in the phases of their respective detection windows. For L- and S-bands, the 
characteristic size of the frequency structure was about 10$-$100\,MHz, varying from pulse to pulse. Two illustrative 
examples of GPs with 100\,MHz and 10\,MHz structure are shown in Fig.~\ref{fig:patches}. These pulses, both P1 GPs,
were recorded during the same L-band observation 20 minutes apart from each other (note that such  a dramatic change in the 
frequency scale of the GP polarization features within a single observing session also  implies that these features are unlikely 
due to calibration errors). For GPs  in the UHF band the width of the patch did not exceed 50\,MHz, although only 11 
sufficiently bright GPs were available for the analysis and a larger number of pulses is needed to confirm any statistical 
 trend.

Emission within a single patch tends to be strongly polarized, with the position angle of linear polarization and the 
hand of the circular polarization varying randomly from patch to patch, and sometimes within a single patch 
(Figs.~\ref{fig:GPs},~\ref{fig:patches}). 
Fig.~\ref{fig:GP_pln} shows  the fractional linear and circular polarizations for both band-integrated GPs and the individual 
1.56\,MHz frequency channels. Here we selected only those band-integrated GPs, or GPs from individual channels, for which 
the S/N  of the total intensity was larger than 15. The calculated fractional polarization incorporates contributions from 
the Stokes components of both the pulsar and the noise.  The Stokes components of the noise fluctuate independently from 
each other around a mean value of 0 with an RMS of 1 (in units of S/N).  Thus, when the pulsar signal has a high level of 
polarization, the absolute value of the calculated fractional polarization can occasionally exceed 1 (Fig.~\ref{fig:GP_pln}). 
Increasing the S/N threshold for total intensity reduced the number of such ``over-polarized'' data points, as expected. 
Overall, band-integrated GPs tend to have smaller fractional polarization, both linear and circular. 

 It must be noted that the observed degree of polarization of individual GPs (both frequency-resolved and band-integrated) 
can be affected by small-number statistics. As was demonstrated by \citet{vanStraten2009}, if the time-bandwidth product of 
recorded pulses is on the order of unity, then the observed degree of polarization does not reflect the intrinsic degree of 
polarization of the source. Although each time sample in our single-pulse data was obtained by averaging 16 (for each frequency 
channel) or 8192 (for band-averaged data) raw voltage samples (see Section~\ref{sec:obs_calib}), the effective number of 
independent samples can be small if the average is dominated by one or a few bright nano-pulses (note that scatter-broadening 
produces correlated samples and thus does not add to the number of degrees of freedom in the average).

 The overall average GPs, obtained by summing all GPs in their respective bands, appear to be unpolarized. 
For L-band, which had the largest number of GP detections, such an average (band-integrated) GP had less than 2\% of both fractional linear
and circular polarizations.

\subsection{Comparison to GPs from other pulsars}

Combining the measurements from this work and from K06a, we can conclude that the observed properties of M28A's GPs, 
such as their short (ns$-\mu$s) duration, high brightness temperature (up to $5\times 10^{37}$\,K), narrow (18\,$\mu$s) 
window of occurrence, power-law energy distribution with the index of about $-1.8$, and the probability of detecting 
GPs with $E>20E_\mathrm{prof}$ in a given period, $P(E>20E_\mathrm{prof})\approx3\times 10^{-5}$ make  the GPs from M28A
rather typical members of the GP population\footnote{The currently known GP emitters (in the sense of  our GP definition 
given at the beginning of the section) are M28A, the Crab pulsar, B1937+21, B1820-30A, B1957+20 and J0218+4232. Strong 
pulses from another young pulsar, B0540-69 are similar to the Crab GPs, but their intrinsic width have not yet been 
identified because of the strong scattering in the ISM \citep{Johnston2003}.} \citep{Knight2005,Knight2006a,Knight2006c}.

It is interesting to compare M28A's GP spectra together with their polarization properties to the spectral structure 
 and polarization of GPs from the most studied GP pulsars, B1937+21 and the Crab pulsar. 

GPs from PSR~B1937+21 have a characteristic width of tens of nanoseconds \citep{Soglasnov2004} and thus are usually unresolved. 
\citet{Popov2003} observed PSR~B1937+21 at 1430\,MHz and 2230\,MHz simultaneously, with  an instrumental fractional 
bandwidth (\frbwi) of 0.01 and 0.02. During the three hours of observing time, spread among multiple sessions, 25 GPs 
were recorded, with none of them occurring at both frequencies simultaneously. This  led the authors to conclude that GPs 
from PSR~B1937+21 have frequency structure with  a scale of \frbwG~$\approx0.5$. In addition, some GPs in their study 
showed variability on  a smaller frequency scale \frbwG~$\approx0.007$, which could not be explained by diffractive 
scintillation. Numerous observations with narrow fractional bandwidths (\frbwi~ranging from 0.001 to 0.03) revealed that 
most of the GPs from PSR~B1937+21 show strong circular and a considerable amount of linear polarization, both varying 
randomly from pulse to pulse \citep{Cognard1996, Popov2004, Kondratiev2007, Zhuravlev2011}. 
\citet{Soglasnov2007} presented an unusual example of a resolved GP, composed of two nano-pulses with different hands of 
circular polarization.  However, all aforementioned works present polarization measurements for individual data samples
(unresolved GPs or unresolved components), thus the reported degree of polarization must be biased by small-number 
statistics \citep{vanStraten2009}.

On  larger frequency scales, GPs from the Crab pulsar exhibit similar  frequency modulation with \frbwG~on the 
order of 0.5 \citep{Popov2008,Popov2009}. On smaller frequency scales, the situation is more complex since it is known
that at frequencies above 4\,GHz, Crab GPs from the main pulse component (MGPs) and the GPs from the interpulse component 
(IGPs) have very different spectral and polarization properties \citep{Hankins2003,Hankins2007}.

MGPs consist of a series of nanosecond-long unresolved pulses (called nanoshots), which often merge together into 
so-called microbursts \citep[see, for example, Fig.~1 in][]{Hankins2003}. Nanoshots are quite narrow-band, with 
\frbwG~$\approx0.03$ \citep{Hankins2007}.  The studies of polarized emission conducted 
by \citet{Hankins2003} and \citet{Jessner2010} %, both with \frbwi~$\approx0.1$, 
showed that individual unresolved
nanoshots are strongly circularly or linearly polarized, with the hand of V changing randomly from shot to shot.
This leads to weak polarization of MGPs if the radio emission is averaged within a pulse \citep{Hankins2007}.
Both below and above 4\,GHz, microbursts from  the Crab's MGPs are known to have broadband 
\citep[\frbwG~between, approximately, 0.3 and 1,][]{Crossley2010}
spectra, with  a scale of fine structure \frbwG~ranging from 0.003 to 0.01 \citep{Popov2008, Popov2009, Jessner2010}.

The GPs from the interpulse do not have nanosecond-scale structure above 4\,GHz
and their spectra are dramatically different from those of MGPs. 
 The IGP spectra consist of bands, proportionally spaced in frequency with $\Delta\nu/\nu = 0.06$ \citep{Hankins2007}. The width 
of the band is 10$-$20\% of the spacing between the bands and sometimes bands can shift upwards in frequency within a 
single GP \citep[see, for example, Fig.~7 in][]{Hankins2007}. \citet{Jessner2010}  studied the polarization smoothed 
within 80\,ns windows (32 samples) for IGPs recorded within \frbwi~$\approx0.06$ and report high degree of linear polarization 
(up to 100\%) and a flat PA curve. Below 4\,GHz there are hints that IGPs have properties similar to those of MGPs 
\citep{Crossley2010,Zhuravlev2013} but more detailed studies are needed.

Due to our coarse time resolution, we can not compare directly the temporal structure of M28A's GPs to that of 
 the Crab's MGPs/IGPs. We do not notice any well-defined, persistent periodicity in the separation between the patches 
in GP spectra, although we must notice that our 100\,MHz modulation has $\Delta\nu/\nu$ of about 0.04$-$0.09, similar 
to  the $\Delta\nu/\nu=0.06$ spacing between the IGP bands. Some of our less modulated spectra could be caused by bands 
drifting up in frequency and smearing the periodic structure. 
 Above 4\,GHz, IGPs of the Crab pulsar have widths of about 3\,$\mu$s \citep{Jessner2010}. If M28A's GPs have similar widths, 
the pulses are resolved and each frequency channel in our data contains several independent samples. 
In this case, the observed degree of polarization may serve as a more or less adequate estimate of the intrinsic GP polarization. 
However, our polarization statistics are hard to compare to that of \citet{Jessner2010} because of the differences in the observing 
setups. We found no published polarization properties of separate frequency bands of Crab IGPs, and thus we do not know if the hand 
of the circular or the PA of linear polarization can vary between the bands of the same GP. 

We find the idea that M28A's GPs consist of individual nanoshots somewhat more compelling. K06a showed an example of 
two strong  pulses from M28A observed simultaneously at 2.7 and 3.5\,GHz. One pulse consisted of a single peak with 
FWHM of 20\,ns, another GP showed multiple unresolved ($<7.8$\,ns) spikes, some of which merged together, forming a 
burst of about 200\,ns. Such behavior is similar to the Crab MGPs. For M28A's GP spectra with well-defined patches, 
the characteristic scale of frequency modulation is close to the spectral width of the Crab's nanoshots, \frbwG~$\approx0.03$. 
The  Crab's individual nanoshots are strongly and randomly polarized and we observe the same for individual patches 
within a single M28A's GP (Fig~\ref{fig:patches}, left). 
Thus, M28A's GPs, with  their well-defined patches, could consist 
of several separate, unresolved, narrow-band nanoshots\footnote{The aforementioned broadband (\frbwG~$\approx0.3$), 20-ns wide GP from K06a does not 
contradict this conclusion, since K06a derive their value of DM for that session by aligning the times of arrival of the 
low- and high-frequency parts for the two GPs recorded. Therefore, it is possible that their 20-ns pulse consisted
of two separate narrow-band nanoshots. } , with a seemingly large apparent degree of polarization. 
 The finer frequency modulation in some of M28A's GP spectra (Fig~\ref{fig:patches}, 
right) has the same order of magnitude as the modulation found in the microbursts of Crab GPs. 
Being composed of  a larger number of overlapping nanoshots would also explain why such GPs have more rapid variation of polarization 
with frequency.

Combining the known spectral and temporal properties of GPs from the PSR~B1937+21, M28A and the Crab pulsar, it is 
tempting to draw a general portrait of a giant pulse. Such a portrait generalizes the information already available, 
but these generalizations must be further tested with a series of dedicated studies. One may speculate that a 
``typical giant pulse'' consists of random number of individual, narrow-band %, strongly but randomly polarized 
nanoshots.  Nanoshots appear to be strongly and randomly polarized, although the intrinsic degree 
of polarization should be constrained by the future studies with much better time resolution.
The number of nanoshots per pulse appears to be bound by some upper limit, and this limit seems to be correlated with the pulsar 
spin period. GPs from PSR~B1937+21, the pulsar with the smallest spin period ($P=1.6$\,ms) among all known GP emitters, 
consist of one or, very rarely, two nanoshots and that is why \citet{Popov2003} failed to detect B1937+21's GPs 
simultaneously at two widely separated frequencies. GPs from M28A ($P=3.05$\,ms) have up to several 
nanoshots, which we have observed as patches in GP spectra. Finally, the Crab pulsar's MGPs 
($P=33$\,ms) are usually made of much larger number of nanoshots which merge together and create continuous spectra. 
The Crab's IGPs, with their lack of temporal structure and banded spectra, clearly do not fit this picture and present 
a separate puzzle. 

\section{M28A in the context of other $\gamma$-ray MSPs}
\label{sec:disc}

\begin{figure*}
   \centering
 \includegraphics[scale=0.95]{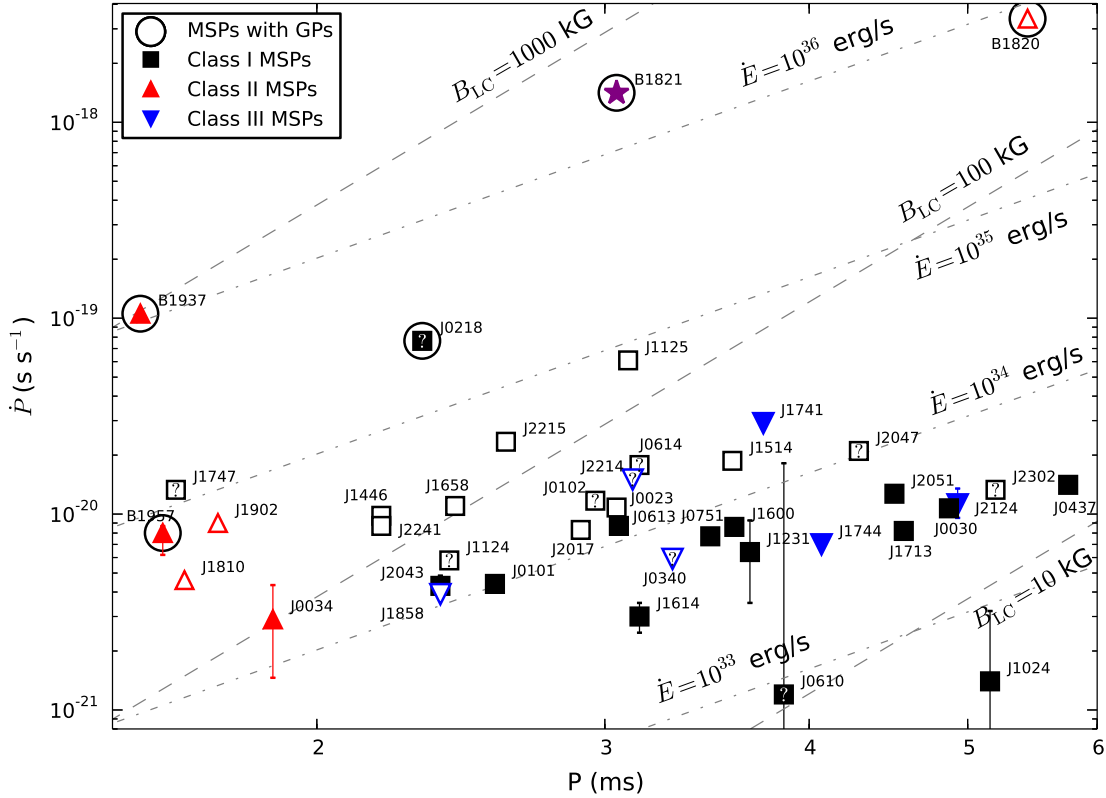}
\caption{M28A (marked with  a purple star) on the $P-\dot{P}$ diagram  of the MSPs from the second Fermi catalog. The names of the 
pulsars are truncated after right ascension and correspond to the full names in Fig.~\ref{fig:indices}. The errorbars 
reflect the estimated uncertainty in $\dot{P}$ due to the Shklovskii effect and  accelerated motion in a gravitational potential.
The pulsars without proper motion measurements are shown with unfilled markers. Following the classification
in J14, Class~I pulsars are shown with black squares, Class~II with red upward triangles and Class~III with the blue downward
triangles. Pulsars which have a different classification in \citet{Ng2014} are featured with a question mark. 
Pulsars with known GPs are  circumscribed with a circle. Dashed lines show constant $B_\mathrm{LC}$ and the dot-dashed 
lines mark constant $\dot{E}$. 
} \label{fig:PPdot}
    \end{figure*}

Here we will review the properties of M28A's multiwavelength emission in the context of tendencies and trends observed 
among the MSPs from the second Fermi catalog. Although M28A, due to its large $\dot{E}$, always was a promising candidate 
to search for  \gr pulsations, the relatively large distance and unfortunate location (both in the Galactic plane and 
in the heart of a globular cluster) made the detection complicated: it took about 190 weeks of Fermi data to detect 
$5.4\sigma$ pulsations (J13). M28A's light curves have not been modeled yet and J13 note that it might be a complex 
case, possibly going beyond the standard models. Thus, a comparison to other, better studied \gr MSPs is important.

According to \citet[][hereafter J14]{Johnson2014}, the MSPs from the second Fermi catalog fall into three categories 
and each category can be more or less well modeled using the simple assumptions about the location of radio/\gr 
emission regions. For the pulsars with \gr peaks trailing the radio peaks \citep[so-called Class~I, after][]{Venter2012}
high-energy emission is believed to come from the broad range of altitudes close to the light cylinder, originating
inside the narrow vacuum gap at the border of the surface of last-closed field lines \citep{Cheng1986, Muslimov2003, Dyks2003}.
Radio emission of Class~I pulsars is thought to come from the open field region above the polar cap and is modeled 
with the core and/or hollow cone components. Both the core and the cone are uniformly illuminated in magnetic azimuth 
(meaning, not patchy) and have the size determined by the statistical radius-to-frequency mapping prescriptions 
\citep{Story2007}. For the Class~II pulsars, the \gr peaks are nearly (within about 0.1  spin periods) aligned with 
the radio peaks, and, unlike Class~I, Class~II radio emission is assumed to originate in the regions significantly 
extended in altitude and co-located with the \gr emission regions \citep{Abdo2010, Venter2012, Guillemot2012}\footnote{Class~II 
pulsars can also be modeled with both \gr and radio emission coming from the small range of heights close to the stellar 
surface inside the narrow gap close to the last open field line \citep{Venter2012}. However, this model results in 
worse fits to the data, according to J14.}. Pulsars with \gr peaks leading the radio peaks are labeled as Class~III. 
In this case,  the radio emission has the same origin as for the Class~I  pulsars, but high-energy emission comes from high 
altitudes above the polar cap volume \citep{Harding2005}\footnote{There exist other models for \gr emission 
\citep{Qiao2004, Du2010, Petri2009}, as well as more detailed (patched cone) models of the radio profiles of MSPs 
\citep{Craig2014}, but we will not focus on them in this work.}.

Figure~\ref{fig:PPdot} shows M28A on the standard $P-\dot{P}$ diagram for the MSPs from the second Fermi catalog. It 
is very important to remember that the observed spin-down rate of MSPs can be greatly influenced by Doppler effects 
 originating from
the pulsar's proper motion \citep{Shklovskii1970} and the acceleration in the local gravitational field. The 
 additional apparent spin-down due to the Shklovskii effect ($\delta\dot{P}_\mathrm{Sh}$) depends on the pulsar period,  the distance 
to the source, and its proper motion. $\delta\dot{P}_\mathrm{Sh}$ is always $>0$ and can be responsible for a large fraction of the 
observed $\dot{P}$, whereas $\delta\dot{P}$ due to acceleration can have either sign, but is usually much smaller than the 
Shklovskii effect \citep[see Table~\,6 in][]{Abdo2013}. The uncertainties in estimating Doppler corrections are 
included in the errorbars\footnote{While most of the proper motion and distance values were taken from the second 
Fermi catalog, for PSR~J0218+4232 we used the proper motion of $6.53\pm 0.08$\,mas~yr$^{-1}$ measured by \citet{Du2014} and 
the distance obtained from the parallax measurements in the same work, but corrected for the Lutz-Kelker bias by 
\citet{Verbiest2014}.} on $\dot{P}$ in Fig.~\ref{fig:PPdot}. For the pulsars with unknown proper motion, the observed 
$\dot{P}$ can be viewed as an upper limit on the intrinsic spin-down rate. Such pulsars are marked with unfilled markers. 
The classes were assigned to pulsars prior to modeling and were mostly based on the lags between their radio and 
$\gamma$-ray peaks. In some cases, the classification differs between J14 and \citet{Ng2014}. In Fig.~\ref{fig:PPdot} 
such pulsars are featured with a question mark inside their markers.

Long before the \gr detection of M28A, \citet{Backer1997}, based on the alignment between radio and X-ray profile, 
put forward the idea that M28A's radio components P1 and P3 may be of caustic origin, whereas P2 comes from above 
the polar cap. Up to now there have been no attempts to model such ``mixed-type'' radio profiles (J14), however, 
previous studies showed that caustic radio emission may have properties somewhat different from the cone/core 
components and that it seems to be limited to the pulsars occupying a specific region on $P-\dot{P}$ diagram 
\citep[][but see discussion in J14]{Espinoza2013,Ng2014}. Below we review the properties of pulsars with both types 
of radio emission and compare them to those of M28A. 

\subsection{Magnetic field at the light cylinder} 

The MSPs with the highest values of $B_\mathrm{LC}$\footnote{The magnetic field at the light cylinder was calculated 
with a classical dipole formula, by assuming that a pulsar is an orthogonal rotator and loses its rotational energy 
only via magnetic dipole radiation. The influence of plasma currents and non-orthogonal magnetospheric geometry may 
alter the values of $B_\mathrm{LC}$ to some poorly known extent \citep{Guillemot2014}.} tend to be Class~II pulsars 
\citep[J14;][]{Espinoza2013, Ng2014}. From Fig.~\ref{fig:PPdot}, we notice that J0034$-$0534 has the 
lowest value of $B_\mathrm{LC}$ among ``clear'' Class~II MSPs, $B_\mathrm{LC}\approx 100$\,kG. However, some of the 
pulsars with $B_\mathrm{LC}<100$\,kG have one of the radio components roughly (within 0.1  spin periods) coinciding 
with a \gr one, for example (but not limited to) PSRs~J0340+4130, J2214+3000 and J2302+4442 (J14). Thus, it is possible 
that caustic radio emission can originate from pulsars with $B_\mathrm{LC}<100$\,kG. Confirmation might be obtained by
direct modeling and/or by studying the profile evolution and polarizational properties of the radio emission from 
these pulsars.

Thirteen MSPs from the second Fermi catalog lie above $B_\mathrm{LC}\approx 100$\,kG line. Out of these pulsars, six 
had been identified as Class~II by J14 and \citet{Ng2014}. Out of these six, three do not have proper motion measurements 
-- J1902$-$5105, J1810+1744 and B1820$-$30A, although for the latter there is indirect evidence that the observed $\dot{P}$ 
is an adequate estimate of the intrinsic pulsar spin-down. PSR~B1820$-$30A is a luminous \gr source and any reasonable 
assumptions about the \gr efficiency leads to the conclusion that most of the observed $\dot{P}$ is intrinsic \citep{Freire2011}. 
The average radio profile of one of these pulsars, PSR~B1957+20, has an additional component which does not match any 
of the \gr peaks and was not modeled by J14.

Two pulsars with Shklovskii-corrected $B_\mathrm{LC}\approx 300$\,kG had been modeled as Class~I in J14: J0218+4232 
and J1747$-$4036. PSR~J0218+4232 has very broad radio profile, with radio emission coming from virtually all spin phases, 
and has a large unpulsed fraction \citep{Navarro1995}. The brighter broad radio component falls at the minimum of the 
\gr profile (which consist of one very broad component), but there is a second, more faint radio  peak with a seemingly 
steeper spectral index \citep{Stairs1999}. This component is not modeled by J14 and it is aligned with the \gr emission. 
\citet{Ng2014} classifies PSR~J0218+4232 as Class~II. PSR~J1747$-$4036 is a newly-discovered \citep{Kerr2012} pulsar 
that has two components in its radio profile, one of which is aligned with the weak \gr peak and the other one is about 
0.4  spin periods behind. \citet{Ng2014} classifies this pulsar as Class~II/I, whereas J14 classifies it as Class~I, 
although they do not model the second radio peak. According to J14, this pulsar has polarization levels which are not 
suitable for a Class~II pulsar, although we will argue below that not all Class~II MSPs have zero linear polarization. 

The last five pulsars above the 100\,kG line are all clear Class~I pulsars, having $B_\mathrm{LC}<150$\,kG and unknown 
proper motions. 

To summarize, none of the well-established $B_\mathrm{LC}>100$\,kG MSPs from the second Fermi catalog have a radio profile 
consisting only of components which have doubtless polar cap (Classes I or III) origin. Such high-$B_\mathrm{LC}$ 
pulsars have either only radio components which are aligned with the \gr profile or the ``mixed-type'' radio profile, 
with some of radio components being aligned and some being not (e.g. PSRs~B1957+20, J1747$-$4036 and perhaps J0218+4232). 
M28A has a very large value of $B_\mathrm{LC}=720$\,kG and in its radio profile two of the three radio peaks roughly 
coincide with the \gr peaks, similarly to PSR~B1957+20. Thus it is not unreasonable to suppose that these two matching 
peaks (components P1 and P3) can be of caustic origin.

It is worth noting that for  young pulsars from the second Fermi catalog\footnote{Young pulsars have much larger 
values of $\dot{P}$ and thus do not suffer from the Shklovskii effect.}, three out of four sources with the highest 
values of $B_\mathrm{LC}$ all have  100\,kG~$<B_\mathrm{LC}<150$\,kG and their radio profiles are clearly misaligned 
with the $\gamma$-rays. The only known young pulsar with aligned profiles is the Crab pulsar, and it has the largest  
$B_\mathrm{LC}$ in the second Fermi catalog, 980\,kG.

\begin{figure}
   \centering
 \includegraphics{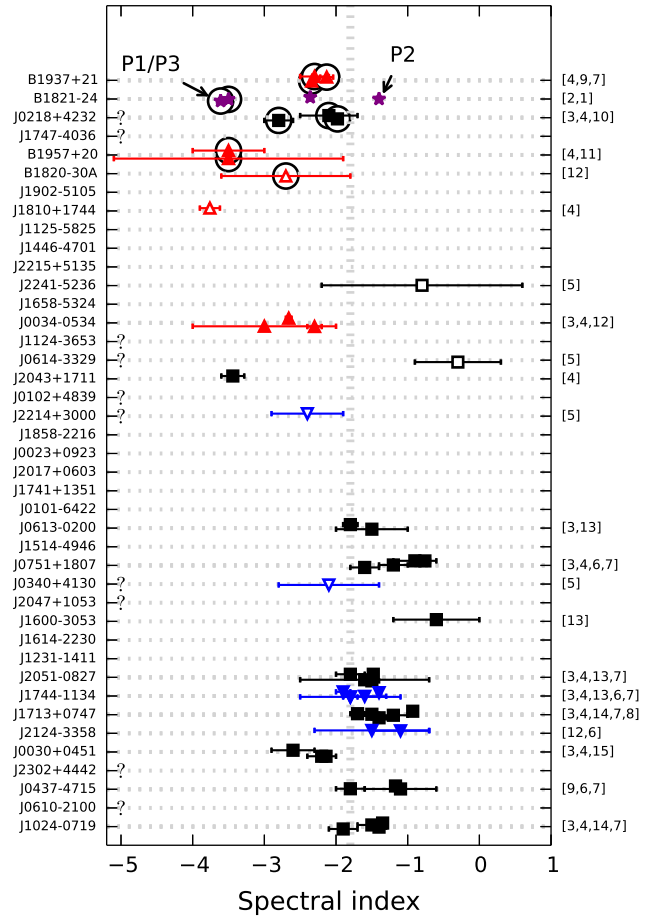}
\caption{All available spectral indices for the radio emission of  MSPs in the second Fermi catalog. 
Pulsars are sorted by descending values of $B_\mathrm{LC}$. The markers follow the same scheme as in 
Fig.~\ref{fig:PPdot}. A small random vertical jitter was added to the markers to facilitate the visual 
examination of the data. A vertical line marks both the mean and the median values of the distribution of MSP 
spectral indices from \citet{Kramer1998}. For M28A, both the spectral index of phase-averaged emission and the indices
from individual components are shown. Note that component P2, the one misaligned with the \gr components
has a shallower spectrum, which is in line with Class~I/III pulsars, whereas components P1/P3 have more steep spectra
and are roughly aligned with the \gr emission.
References: [1] this work; 
[2] PDR14; 
[3] \citet{Kuzmin2001};
[4] Hassall et al in prep;
[5] \citet{Espinoza2013};
[6] \citet{Kramer1998};
[7] \citet{Kramer1999};
[8] \citet{Kijak1998};
[9] \citet{Yan2011b};
[10] \citet{Navarro1995};
[11] \citet{Fruchter1990};
[12] \citet{Toscano1998};
[13] \citet{Ord2004};
[14] \citet{Xilouris1998};
[15] \citet{Lommen2000}.
} \label{fig:indices}
    \end{figure}

\subsection{The spectral index of radio emission} 

According to \citet{Espinoza2013}, Class~II pulsars tend to have steeper spectral indices than the bulk of the \gr MSPs 
(or radio MSPs in general). We compiled a database of spectral indices for the MSPs from the second Fermi catalog and 
 plot them in Fig.~\ref{fig:indices}. The pulsars are arranged by  decreasing $B_\mathrm{LC}$.  
 No spectral information is available for many of them, signifying a good opportunity for future studies. 
Some of the spectral indices 
were measured in different frequency bands, and thus the scatter of points for the sources with multiple measurements 
may reflect an intrinsic departure from a single power-law in the very broad frequency range. However, some of the 
scatter may reflect unaccounted systematic errors (for example, if available spectra consisted of only a few flux 
measurements, with some of the measured fluxes affected by scintillation), 
 and so these index values should be taken with caution, especially those with only one measurement. 

Generally, pulsars with larger values of $B_\mathrm{LC}$, or which are Class~II, have steeper spectral indices than 
the average value for MSPs, $s=-1.8$ \citep{Kramer1998}, whereas Class~I and III have shallower indices, although 
the latter is not true in all cases, e.g. for PSR~J0030+0451.
 
M28A has a quite steep average spectrum, $s=-2.36$ (PDR14), in line with the tendency for the Class~II pulsars. 
Interestingly, components P1 and P3 have much steeper frequency dependence than the component P2 (Table~\ref{table:spind}), 
supporting the idea of different magnetospheric origin. It is worth noticing that the same behavior is exhibited by 
B1957+20, whose shallower component \citep{Fruchter1990} is misaligned with the $\gamma$-rays.

Finally, the Crab pulsar has the spectral index of $-3.1\pm0.2$, unusually steep for a young pulsar \citep{Lorimer1995},
but close to those of the Class~II MSPs.

\subsection{Polarization} 

As was noticed by \citet{Espinoza2013}, pulsars with aligned \gr and radio profiles tend to have a very  small 
amount of linear polarization, which is quite unusual for MSPs. Among six Class~II MSP PSRs from J14, three are 
linearly depolarized. These pulsars are J0034$-$0534, B1820$-$30A and B1957+20 \citep{Stairs1999, Fruchter1990}.
This fact is usually treated as support for the caustic origin of the radio emission, as the radio emission gets 
depolarized coming from different heights \citep{Dyks2004}. However, we must note that the sole presence of 
polarization in the average profile does not preclude a pulsar from being a Class~II. According to \citet{Dyks2004}, 
caustic emission does not  necessitate complete depolarization,  but rather only a decrease in the fractional polarization, 
as well as rapid swings and jumps in the position angle.

Two of the pulsars with aligned radio/high-energy emission, B1937+21 and the Crab pulsar show substantial levels 
of fractional linear polarization: about 0.4 for B1937+21 \citep{Kramer1999} and about 0.2 for the Crab pulsar 
\citep{Moffett1999}. It is puzzling to note that although the positions of the radio peaks for B1937+21 and 
the Crab pulsar are  best reproduced with a caustic radio emission model \citep{Guillemot2012, Harding2008}, 
the PA of the radio emission does not exhibit much variation across the profile components \citep{Ord2004,
Slowikowska2015}. Perhaps more complex models of radio emission are needed, for example the ones which include 
propagation of radio waves through the magnetosphere.

 In M28A, the component P2 shows  a greater degree of linear polarization, and the sweep in PA is reminiscent of that 
predicted by rotating vector model \citep[RVM;][]{Radhakrishnan1969}. Components P1 and P3 show  a much lower amount of fractional  
polarization (although the level is still somewhat higher than for B1937+21 or the Crab pulsar) and not much PA variation, 
except for the seemingly orthogonal jump at the peak of P3. Such PA behavior is similar to  that observed in the main and 
interpulse profile components of both the Crab pulsar and B1937+21.

\subsection{X-ray spectrum and light curves} 

According to \citet{Zavlin2007}, pulsars with large spin-down  energies $\dot{E}>10^{35}$\,erg~s$^{-1}$ and short periods 
$P<3.1$\,ms tend to have non-thermal magnetospheric X-ray emission,  and their X-ray profiles have a large pulsed fraction 
and narrow components. Such X-ray emission  is attributed to synchrotron and/or inverse Compton processes in the 
pulsar magnetosphere. The known MSPs with non-thermal X-ray profiles are M28A, B1937+21 and J0218+4232. For M28A,  the
X-ray components coincide with the trailing edges of radio components P1/P3.

Two other Class~II MSPs, B1820$-$30A and J0034$-$0534 have unfavorable positions  on the sky, with strong X-ray sources nearby
\citep{Migliari2004, Zavlin2006}. PSR~B1957+20 is the prototypical black widow system and a large fraction of its 
X-rays comes from an intrabinary shock between the pulsar wind and that of the companion star \citep{Huang2007}.
Recently, \citet{Guillemot2012} discovered weak, $4\sigma$ X-ray pulsations from this pulsar. The X-ray profile of 
PSR~B1957+20 consists of one broad gaussian-shaped component, which seems to be offset from the \gr components, 
although the authors note that the phase alignment between  the \gr and X-ray data may not be accurate. PSR~J1810+1744 
is another example of an MSP in a black-widow system. The X-ray emission from this pulsar shows broad orbital 
variability, with possible orbit-to-orbit variations \citep{Gentile2014}. So far, there has been no attempt to
search for pulsed X-ray emission from this pulsar. For PSR~J1902$-$5105 only upper limits on its X-ray flux were 
placed \citep{Takahashi2012}.

To our knowledge, no narrow pulsations of clear non-thermal origin have been detected for any other MSP from the 
second Fermi catalog \citep[see Table~16 in][]{Abdo2013}. The rest of the sources have  black-body and/or power-law 
spectra. Broad thermal pulsations have been detected for pulsars J0030+0451, J0751+1807, J2124$-$3358, J0437$-$4715 
and J1024$-$0719 \citep{Webb2004, Zavlin2006, Bogdanov2009}.

\subsection{Giant pulses} 

The idea that GPs are found from pulsars with high values of $B_\mathrm{LC}$ was proposed a long time ago 
\citep{Cognard1996}. Excluding M28A and J0218+4232, all known MSPs that emit GPs are Class~II \gr pulsars (B1937+21, 
B1957+20 and B1820$-$30A, three out of five known MSPs with GPs). For PSR B1957+20, the pulsar with a supposedly 
``mixed-type'' radio profile, GPs have been detected only from one of the Class~II radio components \citep{Knight2006c}. 
It must be noted though, that the statistics of GPs from this pulsar are quite limited, as only 4 GPs were ever
detected \citep{Knight2006c}.

\citet{Knight2005} failed to find GPs from another Class~II pulsar, J0034$-$0534. However, given the amount of the 
observing time in \citet{Knight2005}, their sensitivity limits, and the estimates on the GP rate from the known GP 
MSPs \citep[like, for example PSR~J0218+4232,][]{Knight2006c}, it is possible that \citet{Knight2005} did not have 
enough observing time to detect a sufficiently strong GP with peak flux density above  their sensitivity threshold. Thus, 
further searches of GPs from J0034$-$0534 (as well as from the two other, previously unexplored Class~II pulsars, 
J1902$-$5105 and J1810+1744) could be promising. 

For M28A,  the GPs come from the phase regions on the trailing side of components P1/P3, with component P2 being free 
from any sign of GP emission. 
 If M28A’s components P1/P3 are of caustic origin, then (omitting the complicated case of 
PSR~J0218+4232 and the distant, young, potential GP pulsar B0540$-$69 in the Large Magellanic Cloud) 
we can summarize that all GP pulsars have detected \gr components, whose aligned Class II radio components 
share phase windows with GP emission.
Radio emission from such components is believed to originate in narrow gaps and come frome the 
regions which are both significantly extended in altitude and  positioned near the light cylinder. 
The physical conditions there can be much different from the regions where radio emission is placed traditionally 
( i.e. at low altitudes and within a small range of heights dictated by radius-to-frequency mapping). This should be 
taken into account by the theories of GP generation, for example the ones which rely on plasma turbulence 
\citep{weatherall1998} or on the interaction between the low- and high-frequency radio beams \citep{Petrova2006}.

\smallskip
\smallskip

Based on the discussion above, we can conclude that MSPs with large values of $B_\mathrm{LC}\gtrsim100$\,kG tend to 
have properties different from the bulk of MSPs from the second Fermi catalog. One such property is the presence of 
caustic radio components. It must be noted that the true character of the ``conventional'' and the caustic types of 
radio emission are still to be investigated, since a substantial fraction of MSPs from the second Fermi catalog have 
been discovered just recently and thus have not been extensively studied in radio yet. For M28A, the components P1/P3 
resemble the radio components  typical of Class~II pulsars ( i.e. having a steep spectral index,  rough alignment with 
\gr and X-ray profile components and GP phase windows). Component P2 resembles core/cone emission from altitudes above the 
polar cap ( i.e. having a shallow spectral index, no matching high-energy components, and no GPs). The direct modeling of M28A's 
radio and high-energy profiles promises to be very interesting.

\section{Summary} 
\label{sec:sum}

The extensive set of full-Stokes, wideband and multi-frequency observations of M28A with the Green Bank Telescope 
allowed us to determine a high-fidelity average profile in unprecedented and remarkable detail, as well as to collect 
the largest known sample of M28A's giant pulses.

Between 1100 and 1900\,MHz, we found that M28A's pulsed emission covers more than 85\% of the pulsar's rotation. 
In addition to the three previously described profile components (P1$-$P3), we  distinguish P0
 \citep[first spotted by][]{Yan2011b}, a faint component in the phase window right before the component P1.
We present measurements of phase-resolved spectral indices and 
polarization properties throughout almost all of the on-pulse phase window. The phase-resolved spectral indices 
exhibit prominent variation with spin phase, while at the same time staying roughly similar within four broad phase 
windows corresponding to components P0$-$P3. 

Components P0 and P2 of M28A's profile are almost completely linearly polarized, whereas the levels of polarization 
for P1 and P3 are lower. Two of the four recorded drops in the fractional linear polarization coincide with the 
narrow phase windows of GP generation on the trailing edges of components P1 and P3. The average profile in the 
phase window of GP generation at the trailing edge of component P3 has a small ($15^\circ$) jump of the position 
angle of the linearly polarized emission. Interestingly, in the same phase region, we found an absorption feature 
which resembles a double notch, a feature which may bring some insight into the microphysics of the pulsar radio emission. 

We compare the radio and high-energy properties of M28A to those of MSPs from the second Fermi catalog, and argue 
that M28A's radio emission can be a mix of caustic and polar cap components. The components P1 and P3, characterized 
by steep spectral indices and rough alignment with both high-energy profile components and GP phase windows, resemble 
the caustic radio emission observed previously from PSR~B1937+21, B1820$-$30A and a few other pulsars with a high 
value  of the magnetic field at the radius of the light cylinder. On the other hand, component P2 resembles the 
traditional core/cone emission from altitudes above the polar cap (i.e. with a shallow spectral index, no matching 
high-energy components, no detected GPs). The origin of the P0 component (showing a positive spectral 
index between 1100 and 2400\,MHz, a high level of linear polarization, a rapid PA sweep, and no GPs) remains unclear.

Our measured energy distribution and detection rate for M28A's GPs agrees with previous works. We have also investigated 
the broadband spectra of M28A's GPs. The spectra appear patchy, with the typical size of a patch changing 
randomly from pulse to pulse within the limits of 10$-$100\,MHz.  Individual patches appear to be strongly polarized, with 
the direction of the linear and the hand of the circular polarization changed randomly from patch to patch (and 
sometimes within a single patch). However, depending on the intrinsic GP width (unresolved in our single-pulse data), 
observed degree of polarization may be biased by small-number statistics \citep{vanStraten2009}, 
and thus may not reflect the intrinsic polarization properties of GP emission mechanism. Unlike the main and 
interpulse GPs from the Crab pulsar, M28A's GPs from the trailing edges of components P1 and P3 seem to have similar 
properties to one another. Although our time resolution was not sufficient to resolve the fine temporal structure 
of individual pulses, we argue that M28A's GPs resemble the GPs from the main pulse of the Crab pulsar, which consist 
of a series of narrow-band nanoshots. 

\bigskip 

\begin{acknowledgments} AVB thanks Vladislav Kondratiev, Alice Harding and Jaros\l aw Dyks for the useful discussions, 
Jason Hessels for the comments on the manuscript, Tyrel Johnson for the kindly provided high-energy profiles
 and the anonymous referee for the giant pulse polarization comments. 
TTP acknowledges support from NANOGrav through a National Science Foundation PIRE Grant (0968296).
The National Radio Astronomy Observatory is a facility of the National Science Foundation operated under cooperative 
agreement by Associated Universities, Inc.
\end{acknowledgments}

\bibliographystyle{apj} 
\bibliography{M28A_bibliography} 

\end{document}